# Neural networks for the prediction of peel force for skin adhesive interface using FEM simulation

*Ashish Masarkar[1], Rakesh Gupta[1*], Naga Neehar Dingari[1] and Beena Rai[1]*

[1] TCS Research, SP2 Campus, Rajiv Gandhi Infotech Park, Phase 3, Hinjewadi, Pune

411057, India

*Corresponding Author: gupta.rakesh2@tcs.com, n.dingari@tcs.com

# Abstract

Studying the peeling behaviour of adhesives on skin is vital for advancing biomedical applications such as medical adhesives and transdermal patches. Traditional methods like experimental testing and finite element method (FEM), though considered gold standards, are resource-intensive, computationally expensive and time-consuming, particularly when analysing a wide material parameter space. In this study, we present a neural network-based approach to predict the minimum peel force ($F_{min}$) required for adhesive detachment from skin tissue, limiting the need for repeated FEM simulations and significantly reducing the computational cost. Leveraging a dataset generated from FEM simulations of $90^0$ peel test with varying adhesive and fracture mechanics parameters, our neural network model achieved high accuracy, validated through rigorous 5-fold cross-validation. The final architecture was able to predict a wide variety of skin-adhesive peeling behaviour, exhibiting a mean squared error (MSE) of $3.66 \times 10^{-7}$ and a $R^2$ score of 0.94 on test set, demonstrating robust performance. This work introduces a reliable, computationally efficient method for predicting adhesive behaviour, significantly reducing simulation time while maintaining accuracy. This integration of machine learning with high-fidelity biomechanical simulations enables efficient design and optimization of skin-adhesive systems, providing a scalable framework for future research in computational dermato-mechanics and bio-adhesive material design.

**Keywords:** *Peel force, neural networks, skin-adhesive interface, FEM, cohesive zone model*

# Introduction

The peeling behaviour of adhesives on biological tissues, such as human skin, is a critical aspect of biomedical and industrial applications, including adhesive tapes [1], wound dressings [2], wearable sensors, implantable devices [3,4,5] and tissue repair and regeneration [6]. Understanding and predicting the interactions between adhesives and the skin under peeling conditions are essential to ensure optimal performance and minimize adverse effects such as skin irritation or damage [7,8]. These interactions are governed by the complex interplay of mechanical properties of the skin and adhesive materials, as well as the fracture mechanics at the skin-adhesive interface. Several theoretical models have been developed to describe the peeling behaviour between these two distinct surfaces, with Rivlin's model being one of the most widely recognized [9]. To capture the effects of viscous processes [10], complex loading

[11], and environmental variations on surface adhesion [12,13], researchers have extended Rivlin's model by incorporating parameters such as peeling velocity, preload, and the mechanical properties of the backing and substrate, thereby improving its accuracy in characterizing adhesive interactions [10–16].

Traditionally, finite element method (FEM) simulations have been employed to model and predict peeling behaviour using the cohesive zone modelling (CZM) due to their robustness in capturing intricate material behaviours and interfacial mechanics [17]. A recent study demonstrated the consistency between the experimentally-derived and computationally-estimated peel forces across various peel angles, testing rates, and pressure sensitive adhesive (PSA) layer thicknesses using FEM [18,19]. Similarly, Nase et. al. analysed the peel behaviour of polymer films both experimentally, and through FEM simulations, highlighting an exponential decrease in peel force with increase in the additive content and higher peel rates [20]. However, majority of these investigations primarily concentrated on modelling the peeling dynamics with polymeric substrates like polyethylene or with thin films, thereby neglecting the fundamental biomechanical interactions at the skin-adhesive interface. More recently there has been studies focusing on modelling the adhesive performance of thin films or tapes on rigid and tough substrates, where FEM simulation has been employed to evaluate the peeling process of thin films to analyse the effects of bending stiffness, adhesion energy and heterogeneity in the film [21]. In another study by Garg et. al. the peeling behaviour of heterogeneous thin films on rigid substrates has been examined via experimental techniques and finite element analysis indicating that the peak peel force for homogeneous thin films is observed prior to steady-state peeling, affected by bending stiffness and adhesion energy at the interface [22]. Yin et. al. showed that the results from the FEM simulations of the peeling process on rigid substrates were in quantitative agreement with the theoretical predictions, revealing that peeling force is influenced by parameters such as interface strength and toughness before reaching a steady state [23].

Several other studies have been conducted to characterize the peeling behaviour of various films adhered to rigid substrates [24,25,26]. However, most of the FEM simulation studies of peeling adhesives have predominantly concentrated on interactions with either soft or rigid substrates, providing insights into mechanics of adhesion in these simplified contexts. The intricacies of skin tissue, which presents a unique combination of mechanical properties, biological complexities, and dynamic responses, have largely been overlooked in existing studies [27]. Unlike conventional substrates used for quantifying the adhesive performance, skin is a highly nonlinear, anisotropic, and viscoelastic tissue that significantly affects the adhesive performance [28]. Adhesive performance on the skin varies with physicochemical properties which in turn varies with age, gender, ethnicity, health and environment, posing challenges for designing universally effective formulations [29]. Moreover, the substrates used for adhesion testing based on industrial standards possess surface properties distinct from those of human skin. As per our information, there has been no study focusing on modelling the peeling performance of adhesives on skin as a standardized substrate that produce data predictive of adhesion to skin over time [30]. Although Yang et. al. showed that there is significant variation in the adhesion efficacy of tough adhesive hydrogels across diverse tissue substrates [31]. But the methods used in the study rely heavily on experiments for parameter estimation and FEM simulations which are computationally expensive, time-intensive, and require expert intervention for tuning and interpretation, making them impractical for high-throughput or real-time applications [32].

To address these challenges, we propose methodology that integrates FEM simulations with machine learning to enable rapid and accurate prediction of peeling behaviour. Several authors have noted that machine learning techniques, such as artificial neural networks, can reduce computational costs while maintaining the accuracy of numerical methods [33,34,35,36]. At first, we have generated a dataset of force-displacement curves using the FEM simulations by systematically varying adhesive material properties and fracture mechanics parameters within ranges established from the literature. This dataset was then used to train a neural network capable of predicting the minimum peel force for given input parameters, thereby eliminating the need for additional FEM simulations. The neural network not only significantly reduces computational cost but also provides a practical tool for optimizing adhesive designs and evaluating new formulations.

In this work, we present the detailed methodology for generating the FEM dataset, training the neural network, and validating its performance. The results highlight the feasibility of mimicking the computationally expensive FEM simulations with a machine learning model, paving the way for future advancements in adhesive design and peeling mechanics' research. This study not only provides a computationally efficient solution but also contributes to the growing body of research at the intersection of biomechanics, fracture mechanics, and artificial intelligence.

## Methods

The methodology of this work involves generating the force-displacement (F-d) data for the peeling process of adhesive with skin as a substrate, performing the FEM simulations, and then training a neural network model on this dataset. The details about each of these tasks is described in the following subsections. Figure 1 illustrates the overview of our entire workflow starting from simulation to the model evaluation.

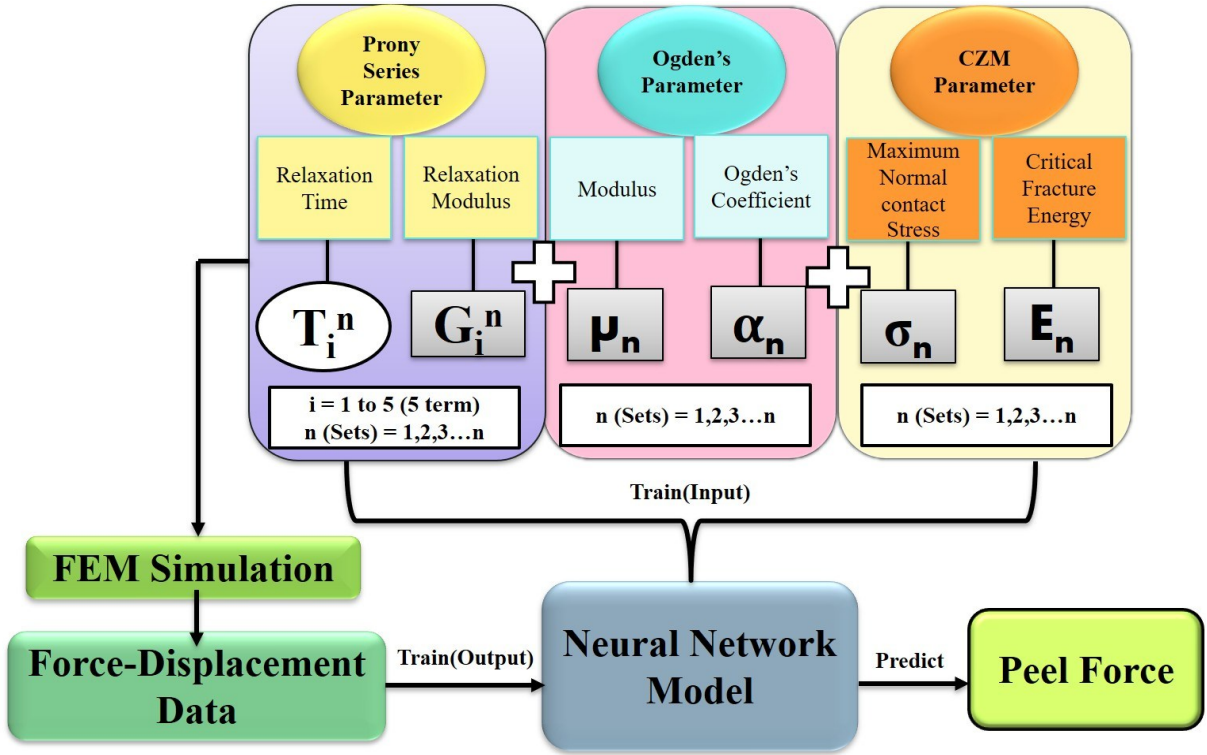

Figure 1: Workflow of the present study. Material model and cohesive zone parameters are varied in combination to generate FEM simulation data of the $90^0$-peel test which are fed as input and output for the neural network model for predicting the minimum peel force.

## **Peel Test**

The $90^0$-peel test is a widely used method for quantifying the adhesion performance of medical adhesives, providing critical insights into the material's ability to resist separation under applied force. During the test, a substrate is bonded to the adhesive with some overlapping region, and the free end is peeled away in vertical direction at a constant rate, generating a peel force-displacement (F-d) curve that reflects the adhesive's mechanical properties. The F-d curve obtained from a peel test is an essential tool for characterizing the adhesive behaviour of materials, as it delineates the relationship between the applied force and the resultant displacement during the peeling process [38]. The curve typically begins with a debonding phase, where the adhesives starts to separate from the substrate, accompanied by a sharp increase in force due to the initial resistance of the adhesive bond. As peeling progresses, the force fluctuate due to irregularities in the adhesive or the surface, but eventually the curve reaches a steady state region, characterized by a relatively constant force required to continue peeling [39]. The minimum peel force, often observed at the onset of debonding or during the transition to steady-state, is a critical parameter because it indicates the initial energy required to break the adhesive bond, directly influencing the adhesive's performance under practical conditions [40]. The peel work of adhesion ($W_a$) is taken to be equivalent to critical energy release rate ($G_a$) which is calculated as area under the positive region of the F-d graph also known as fracture energy with the metric units J/m$^2$ (also expressed as force per unit width N/m, common in peel test reporting) [41]. Maximum normal contact stress ($\sigma_{max}$) is measured directly as the peak stress in kPa of the same graph.

## Visco-hyperelastic Material Model

Adhesives exhibit visco-hyperelastic behaviour, which requires a combination of hyperelastic and viscoelastic modelling framework. A common approach to model viscoelasticity in adhesives is the use of Prony series parameters, often coupled with a hyperelastic framework such as the Ogden model [42]. The constitutive behaviour of the adhesives under step strain relaxation is characterized by both strain and time dependence. The most commonly used mathematical representation of this behaviour is given by Prony series as [43]:

$$g_R(t) = 1 - \sum_{i=1}^{N} g_i \left(1 - e^{-\frac{t}{\tau_i}}\right) \tag{1}$$

Where $g_i$ is a material constant and $\tau_i$ is the relaxation time. $g_R$ is the dimensionless shear relaxation modulus given as:

$$g_R(t) = \frac{G(t)}{G_o} \tag{2}$$

where $G(t)$ and $G_o$ are the time dependent and the initial shear modulus, respectively. For hyperelastic material models, the relaxation equation is typically applied to the constants that define the energy function.

In this study, we have used Ogden's hyperelastic material model, to account for the hyperelastic behaviour of the adhesive, in addition to the viscoelastic nature. Similarly, for the substrate, Ogden's material model is used which more accurately describes the hyperelastic behaviour of the skin tissue as studied in the literature [44]. The strain density function for Ogden's model can be expressed in terms of principal stretch ratio as:

$$W = \sum_{i=1}^{N} \frac{2\mu_i}{\alpha_i^2} (\lambda_1^{\alpha_i} + \lambda_2^{\alpha_i} + \lambda_3^{\alpha_i} - 3) \tag{3}$$

where $\lambda_i$, $i = 1,2,3$ is the $i^{th}$ principal stretch; μ is the shear modulus and α is the Ogden coefficient. For simplicity, we have taken first order Ogden's model (i.e. N = 1) for both the adhesive and the skin tissue.

## Cohesive Zone Modelling

The interfacial bonding between the adhesive and the skin substrate is modelled using bilinear triangular cohesive zone model, a framework that has been extensively employed to investigate the interaction between interfacial properties and bulk deformation in various kinds of materials [45,46]. More recently, this model has been successfully adapted to describe cohesive failure in soft materials [47,48]. Figure 2 illustrates the peel test geometry and the bilinear triangular cohesive zone model implemented for defining the fracture mechanics used in the simulation. The critical fracture energy, G, is defined in terms of $\sigma_{max}$ and the maximum separation distance ($\delta_{max}$) given by the following relationship:

$$G = \frac{1}{2} \cdot \sigma_{max} \cdot \delta_{max} \tag{4}$$

The value of the parameter $G$ is characterized experimentally using fatigue tests [49]. The accurate quantification of $\sigma_{max}$, presents significant challenges that complicates its precise measurement [31]. The value of $\sigma_{max}$ for porcine skin usually falls in the range of 10 KPa to 100 KPa, typically treated and varied as a fitting parameter [50,51]. This range is carefully selected based on typical adhesive strength values reported for various tissue adhesives in lap-shear tests, as documented in the literature [52]. Therefore, we have also varied the value of $\sigma_{max}$ around the same range, in order to generate different F-d dataset to be used for further model building.

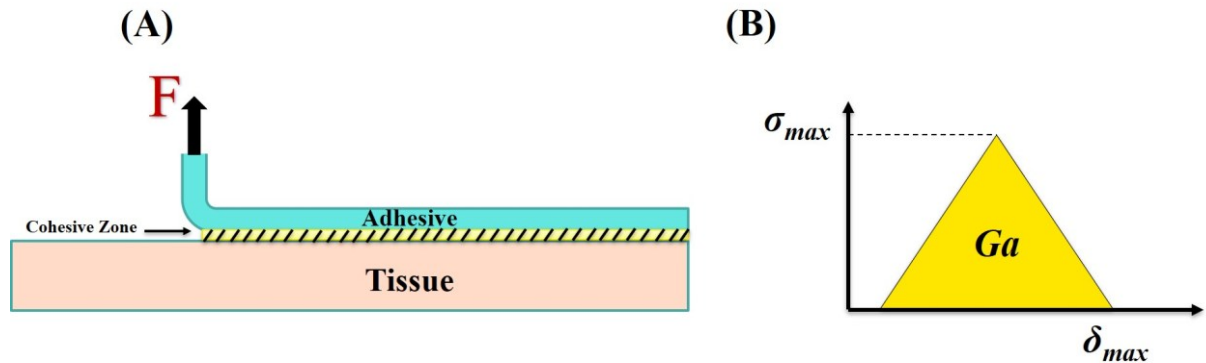

Figure 2: (A) $90^0$ peeling of the adhesive from skin tissue as a substrate with cohesive zone at the interface (B) Interfacial peeling is modelled using a bilinear triangular cohesive zone model.

## Finite Element Simulation

A two-dimensional (2D) plain strain finite element model of the 90º peel test was developed using the FEM software ANSYS Workbench 2023. The thickness of the adhesive layer was taken to be 0.15 mm because it falls within the optimum value range for adhesives thickness according to the literature [37]. The length of the skin substrate was assumed to be 80 mm with the overlapping region between the adhesive layer and the substrate as 40 mm. The thickness of the skin tissue substrate was taken as 3 mm which was found to be sufficient to mitigate the peeling process [31]. A bonded surface body to surface body contact was used as the scoping method for the interface between the adhesive layer and substrate. A triangular traction separation law was employed with two parameters $\sigma_{max}$ and $G_a$ defining the CZM with the contact debonding module. The free end of the peeling arm was displaced at the rate of 5mm/sec, in vertical direction for the $90^0$-peeling test, while the substrate was restrained from both the horizontal and vertical direction. The schematic of the peel test geometry created in design modeler with adhesive layer and skin tissue as a substrate has been shown in the Figure S1 (A). More details about the FEM simulation has been given in supplementary information SI.

The F-d data was generated by systematically varying the Prony series and Ogden's model's parameters (equation 1-3) of the adhesive. Specifically, different combinations of relaxation times and moduli from the Prony series were tested alongside varying Ogden parameters i.e. the shear modulus µ and the Ogden's coefficient α. Along with that the CZM parameters were also varied to capture a wide range of interfacial behaviours between the adhesive and the substrate (equation 4). Skin substrate was modelled as hyperelastic material with Ogden's hyperelastic material parameters being constant for each FEM simulation.

This systematic variation enabled the generation of a comprehensive dataset that reflects a wide range of adhesive behaviours. The details about different combinations of material model parameters and CZM parameters used in the FEM simulation has been given in supplementary table S1. For each material and CZM parameters, F-d data was obtained using FEM simulation. The minimum peel force ($F_{min}$) required for the debonding of the adhesive and substrate was calculated. The aim of this study is the prediction of $F_{min}$, given the adhesives material and fracture mechanics parameters using machine learning techniques.

Remark 1: It is also worth to mention that the computational time taken for each full finite element run takes between 20 - 40 minutes depending on the material parameters and the number of mesh elements and solver configuration.

# Machine Learning Implementation

To predict the minimum peel force required for the interface peeling of adhesive from skin, a robust and systematic machine learning framework was developed. The proposed methodology involved determining the optimal architecture for a feedforward neural network using an exhaustive hyperparameter search followed by rigorous evaluation through K-fold cross-validation. This multi-step approach ensured that the final model was both well-tuned and generalizable, despite the limited dataset derived from FEM simulations.

## Data Preparation and Preprocessing

The dataset, consisting of material model parameters and FEM generated peel force data, was preprocessed to ensure compatibility with neural network training. The FEM simulation data generated using Ansys consisted of 18 features including Ogden's material model parameters for both skin and adhesive, 5 term Prony series parameters for adhesive as well as cohesive zone parameters describing the fracture mechanics. The output variable of the simulation, representing the minimum peel force i.e. $F_{min}$ was selected as the target variable. Since skin substrate was kept as constant for each simulation run, the Ogden's material parameters $\mu_{skin}$, $\alpha_{skin}$ and incompressibility parameter '$d_{skin}$' of skin were excluded from dataset for the model building. Table 1 summarizes all the features used for model building along with their descriptions.

Features were standardized using z-score normalization to enhance convergence during optimization, while target variable was similarly scaled for numerical stability. The dataset was divided into training and test sets using a 90%-10% split, ensuring a random distribution to prevent overfitting. Data preprocessing was performed using Scikit-learn library. The K-fold class from Scikit-learn library was utilized to implement a 5-fold cross-validation approach. This method partitions the training dataset into five mutually exclusive folds. During each iteration, four folds are used for training the model, while the remaining fold is held out for validation. By rotating the validation fold across iterations, this strategy ensures that every data point contributes equally to model training and evaluation, thereby maximizing the utility of the limited dataset. The number of folds, k=5 was chosen to balance computational efficiency and a reliable estimate of model performance. Shuffling the data before splitting was enabled, with a fixed random seed (random state = 42) to ensure reproducibility of results.

Table 1: Input and the output features

| FEM simulation input parameters used for the neural network model | |
|---|---|
| $\mu_{adhesive}$ | Shear modulus of adhesive for Ogden's hyperelastic model |
| $\alpha_{adhesive}$ | Ogden's coefficient of the adhesive |
| d | Incompressibility parameter of the adhesive |
| $g_i$ (i = 1-5) | Relaxation modulus of the adhesive for 5 term Prony series |
| $t_i$ (i=1-5) | Relaxation times of the adhesive for 5 term Prony series |
| $\sigma_{max}$ | Maximum normal contact stress for bilinear triangular cohesive zone model |
| G | Critical fracture energy |
| FEM-generated output feature of the neural network model | |
| $F_{min}$ | Minimum peel force required for the interfacial peeling between skin and the adhesive |

## **Architecture Selection Using Cross-Validation**

We employed a grid search strategy to determine the optimal neural network architecture. The hyperparameter space included:

- Number of hidden layers: Models with 1,2 and 3 hidden layers were tested, enabling evaluation of shallow versus deeper architectures.
- Number of neurons per layer: Within each layer, 4,6,8,16,32 and 64 neurons were considered to investigate the impact of model complexity.
- Regularization (L2 penalty): Fixed at 0.01
- Dropout rate: Set as 20% during hyperparameter tuning for robustness.

For each combination of hidden layers and neurons, the following steps were performed:

1. Data Partitioning: Each fold of training and validation data was defined by the indices generated by the Kfold splitter.
   - $X\_{train\_fold}$ and $y\_{train\_fold}$: Data used for training in the current fold.
   - $X\_{val\_fold}$ and $y\_{val\_fold}$: Data reserved for validation in the current fold.
2. Model Initialization: A custom function was used to instantiate a neural network with the specified number of layers and neurons. The model included input layer dimension matching the input features X, several hidden layers and an output layer. Fully connected hidden layers, each with rectified linear unit (ReLU) activation function was constructed. Additionally, L2 regularization was applied on the weights to prevent overfitting and enhance the generalization ability of the model. Dropout layers were also included to further mitigate the overfitting by randomly setting a fraction of the input units to zero during training. The output layer consisted of a single neuron for predicting the target variable $F_{min}$.

3. Model Training: The model was trained using the ADAM optimizer with mean squared error (MSE) as the loss function. The training was conducted over 500 epochs with a batch size of 8, and validation performance was monitored after each epoch.
4. Validation: Predictions on the validation fold were scaled back to the original range using the inverse transform of the target scaler. MSE, mean absolute error (MAE) and $R^2$ score metrics were used for the evaluation of the model's performance in each fold.
5. Metric Aggregation: For each architecture, the average MSE, MAE and $R^2$ scores were computed for each fold (equation 5-7) (where $K = 5$), and the average values were used to rank model configurations.

$$Average\ MSE = \frac{1}{k}\sum_{i=1}^{k} MSE_i \tag{5}$$

$$Average\ MAE = \frac{1}{k}\sum_{i=1}^{k} MAE_i \tag{6}$$

$$Average\ R^2 = \frac{1}{k}\sum_{i=1}^{k} R_i^2 \tag{7}$$

## Model Development

The final model consisted of a fully connected neural network model, having the best combination of the number of neurons per layer and the number of hidden layers, identified during the hyperparameter optimization using 5-fold cross validation method. The input layer had 18 features, a hidden layer with 32 hidden units and one output layer for the target variable. The input features were fed to the network in batches of 8. Adam optimizer with default learning rate and MSE loss function was used to train the model for each fold. To further enhance the model's robustness, an early stopping mechanism was employed, monitoring the validation loss and terminating training if no improvement was observed over 10 consecutive epochs. This approach prevents overfitting while ensuring sufficient training. During the training process of 1000 epochs, model with the best performance was saved. Table S3 summarizes the hyperparameters used for the final neural networks model.

A second round of 5-fold cross validation was conducted on the entire training dataset using the finalized model architecture. The additional evaluation step ensured that the selected configuration consistently outperformed alternatives and was not a result of random fluctuations in data portioning. All the models were implemented using the 'Keras' API of 'TensorFlow'. More detailed explanation about the feedforward neural networks model used in this study is given in supplementary information.

# Result and Discussion

# Optimizing Architecture

The performance of the neural network model for predicting $F_{min}$, to detach the adhesive from the skin was evaluated using a 5-fold cross-validation technique, with variations in the number of hidden layers and neurons. The results, summarized in terms of MSE, MAE, and $R^2$ score, provide insights into the optimal network architecture for this task. The first set of experiments focused on exploring different network architectures with varying numbers of hidden layers and neurons. For a single hidden layer, the model's performance improved as the number of neurons increased. The architecture with one hidden layer and 32 neurons yielded the best results with the lowest MSE ($5.55 \times 10^{-7}$), the smallest MAE (0.000392), and the highest $R^2$ score (0.82) as shown in Figure 2. The low MSE indicates that the model's predictions are very close to the true values, while the small MAE suggests minimal average deviation between the predicted and actual minimum peel forces. The $R^2$ score of 0.82 demonstrates that approximately 82% of the variance in the minimum peel force can be explained by the model, which is a strong indicator of model performance.

Interestingly, the performance of the model with one hidden layer continued to improve as the number of neurons increased. With 4 neurons, the model showed a relatively higher MSE ($1.55 \times 10^{-6}$) and a lower $R^2$ score (0.50), indicating that the model was underfitting the data. As the number of neurons increased to 8 and 16, the performance improved, with $R^2$ scores rising to 0.71 and 0.73, respectively. This improvement demonstrates the model's capacity to better capture the relationships between the input features and the target variable ($F_{min}$) as more neurons were added, allowing for more complex representations of the data. The architecture with 32 neurons achieved the optimal balance, yielding the best overall performance. When two hidden layers were introduced, the results did not significantly outperform the single-layer configuration. From Figure 2A, the architecture with two hidden layers and 32 neurons achieved a slightly higher MSE ($8.47 \times 10^{-7}$) compared to the single-layer 32-neuron model, and the $R^2$ score (0.71) was marginally lower as can be seen in Figure 2C. Similarly, increasing the number of neurons within each layer did not produce notable improvements. These findings suggest that the addition of more layers did not contribute to a substantial increase in model performance, potentially due to the relative simplicity of the relationship between the input features and the target variable. Similarly, when three hidden layers were explored, the model performance showed only a slight improvement over the two-layer configurations. For instance, the architecture with three hidden layers and 32 neurons achieved a slightly lower MSE ($8.88 \times 10^{-7}$) and a slightly higher MAE (0.000478) compared to the two-layer 32-neuron model. The $R^2$ score (0.69) was also lower than that of the one-layer 32-neuron configuration. This suggests that while adding complexity through more hidden layers can capture more intricate patterns in the data, it does not necessarily result in better predictive performance in this case. The results from varying the number of hidden layers and neurons clearly indicate that increasing the number of neurons in a single hidden layer enhances the model's performance. However, beyond a certain point, additional layers did not improve the model's ability to predict the minimum peel force accurately. Figure 2C illustrates the comparison of $R^2$ score for different combinations of hidden layers and the number of neurons. The best performance was achieved with a single hidden layer and 32 neurons, which appears to be the optimal network architecture for this task suggesting a simpler relationship between the input parameters and target, rather than a complex one.

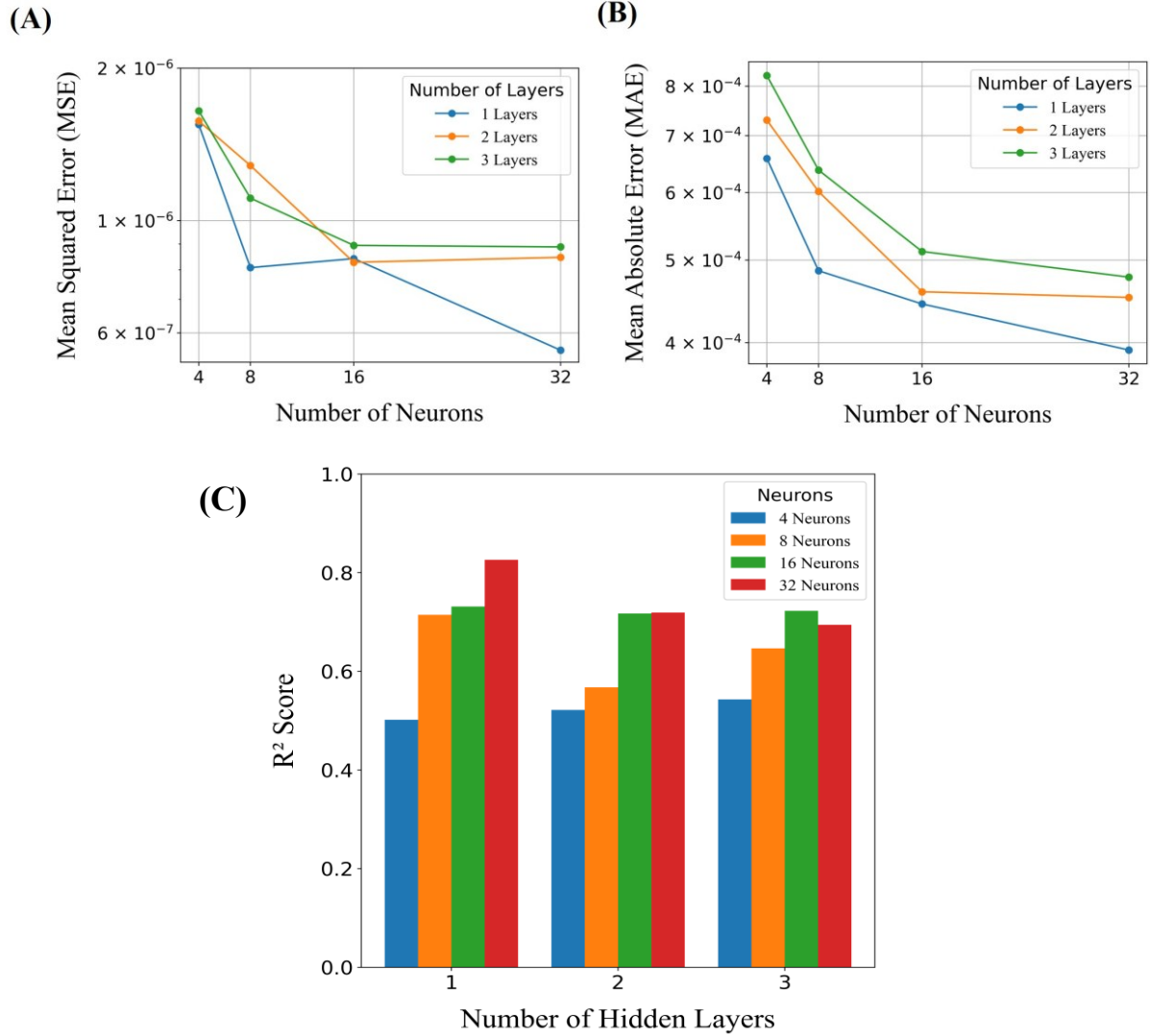

Figure 3: (A) Average Mean Square Error (MSE) (B) Average Mean Absolute Error (MAE) (C) Average coefficient of determination 'R$^2$' score comparison from 5-fold cross validation with the number of neurons for varying hidden layer.

## **Model Performance**

After determining the optimal architecture for the neural network, a final model with one hidden layer comprising 32 neurons was implemented to predict $F_{min}$ based on the input parameters. The training and validation loss progression over the epochs in Figure 4 indicates the model's learning behaviour. The consistent decrease of the loss curves over the epochs reflects the effectiveness of the neural network model in learning the relationship between the input material and fracture mechanics parameters with the predicted peel forces. At the initial epoch, the training loss was recorded at 3.56, with the validation loss slightly lower at 2.46. This small initial disparity between training and validation loss suggests the model was not overfitting during the early stages of training. As the epochs progressed, the training and validation loss steadily decreased, signifying the network's ability to generalize well on unseen data. By approximately the 50th epoch, the validation loss starts to stabilize, reflecting that the model has captured the fundamental patterns in the dataset.

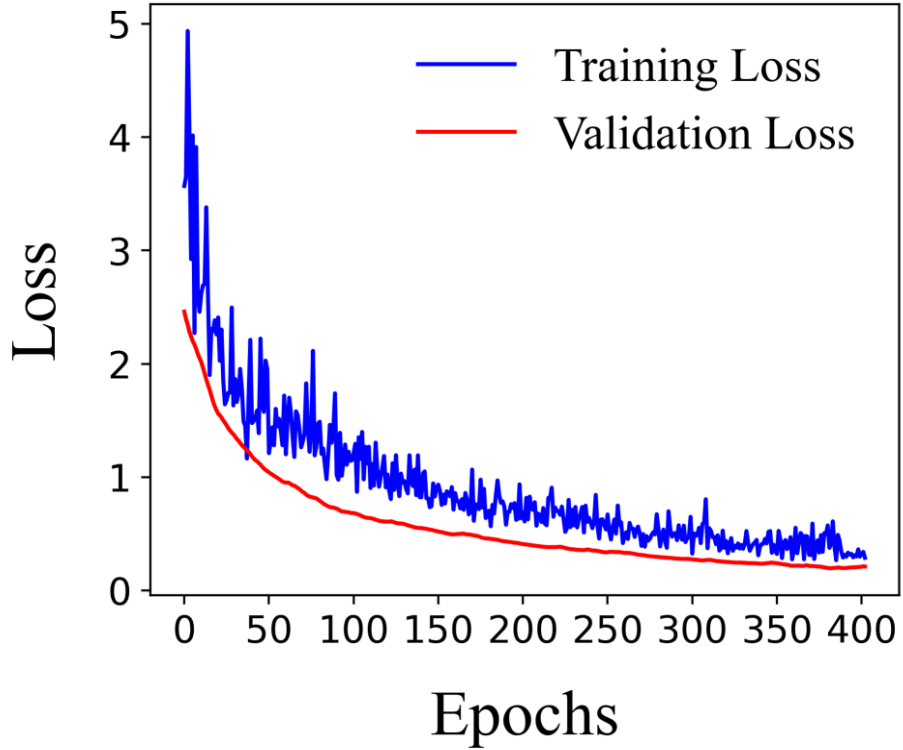

Figure 4: Training and validation curve loss for the final neural network model with optimized network architecture.

The validation loss remains lower than the training loss throughout the training process, indicating that the model was not overfitting. This behaviour can be attributed to the inclusion of regularization techniques such as dropout, as well as strategies like early stopping, which prevent the model from excessively fitting to the training data. The final validation loss demonstrates minimal fluctuation, further affirming the model's stability and robustness in predicting unseen data. The training loss exhibits small oscillations even after 100 epochs, which can be attributed to the stochastic nature of gradient descent. These fluctuations decrease as the number of epochs increases, with the training curve converging steadily. Such noise in the training loss is typical since the neural network model was trained on relatively small datasets and the gradients may vary due to batch sampling.

By the end of 400 epochs, the training loss achieves a minimal value with minimal oscillations, reflecting that the model has adequately minimized the error on the training data. The validation loss reaches a near-constant value, indicating that the model had reached its optimal capacity to fit the training data while minimizing error on unseen validation data This balance between training and validation losses demonstrates the efficacy of the model for predicting the minimum peel force (Fmin) based on input parameters. The consistent performance can be attributed to the balanced architecture of the neural network, which integrates regularization techniques, an appropriate number of hidden layers, and well-tuned hyperparameters.

One of the most significant outcomes of this study is the ability to predict the peeling behaviour of various adhesives using only material parameters and fracture mechanics parameters. This capability eliminates the need for repetitive and computationally intensive FEM simulations or experimental testing. By leveraging these parameters as inputs, the model is able to provide insights into the peeling behaviour of different adhesive formulations, making it a highly efficient and cost-effective tool for skin adhesion testing. Unlike traditional FEM-based approaches, which are computationally intensive, the neural network achieves accurate predictions with significantly reduced computational overhead. The observed rapid convergence of the loss curves within a few hundred epochs further underscores the efficiency of this machine-learning-driven approach. Table summarizes the key performance metrics of the final model on the training, validation and test sets. The final model achieved low MSE and MAE values across all dataset, with high $R^2$ scores indicating the proportion of variance in the data explained by the model.

Table 2: Mean Square Error (MSE), Mean Absolute Error (MAE) and R2 score of the final neural network model for the training, validation and test sets

| Performance Metrics | Training | Validation | Test |
|---|---|---|---|
| MSE | $2.18 \times 10^{-7}$ | $3.22 \times 10^{-8}$ | $3.65 \times 10^{-7}$ |
| MAE | $2.79 \times 10^{-4}$ | $1.29 \times 10^{-4}$ | $4.43 \times 10^{-4}$ |
| $R^2$ | 0.93 | 0.99 | 0.94 |

The training data achieved a very low MSE of $2.18 \times 10^{-7}$, a MAE of $2.53 \times 10^{-4}$ and an excellent $R^2$ score of 0.93 indicating a good fit to the training data. The validation metrics were closely aligned with the training performance, demonstrating that the model generalize well on unseen data during training, maintaining consistency between training and validation datasets. The performance of the test set confirms the robustness and reliability of the final model with an MSE of $3.65 \times 10^{-7}$, a MAE of $4.43 \times 10^{-4}$, and an $R^2$ score of 0.94 suggesting that the model retains its predictive power even when applied to entirely unseen simulations. As far as the authors are aware, this study introduces the most accurate comprehensive framework integrating FEM simulation with machine learning for predicting the peel forces in skin adhesive peeling models. However, it is important to note that a direct comparison with traditional methods is not feasible, as this type of peeling study for skin-adhesive systems, particularly focusing on minimum peel force prediction using FEM simulations, has not been comprehensively explored before.

## Cross Validation

When working with small datasets, obtaining reliable and unbiased evaluation metrics for machine learning models can be challenging. Therefore, for this study, we employed 5-fold cross-validation which provides a more robust and accurate approach to assess model performance compared to a simple train-test split. The cross validation applied on the final neural network model exhibited promising performance across all five folds. Table 3 summarizes the 5-fold cross validation performance metrics across each fold. The average MSE ($4.35 \times 10^{-7}$) and MAE ($3.37 \times 10^{-4}$) were consistently low, suggesting that the model was capable of minimizing the difference between the predicted and actual peel force values. The $R^2$ scores averaged across the folds were also high with a final averaged score of 0.90 signifying a strong correlation between the predicted and observed outcomes.

Table 3: 5-fold cross validation MSE, MAE and R2 score across each split.

| Folds | MSE | MAE | $R^2$ |
|---|---|---|---|
| 1 | $1.48 \times 10^{-6}$ | $7.02 \times 10^{-4}$ | 0.67 |
| 2 | $1.06 \times 10^{-7}$ | $2.25 \times 10^{-4}$ | 0.95 |
| 3 | $5.04 \times 10^{-7}$ | $4.95 \times 10^{-4}$ | 0.90 |
| 4 | $6.93 \times 10^{-8}$ | $1.72 \times 10^{-4}$ | 0.99 |
| 5 | $1.62 \times 10^{-8}$ | $9.35 \times 10^{-5}$ | 0.99 |
| Averaged Metrics | $4.35 \times 10^{-7}$ | $3.37 \times 10^{-4}$ | 0.90 |

The training and validation loss curves across the five folds shown in Figure 5 provide a detailed insight into the neural network model's learning behaviour and its generalizability across varying data splits. These are critical in evaluating the robustness of the model, particularly when the dataset is small, as in this case. Each fold reflects distinct characteristics in convergence behaviour, stabilization trends, and potential overfitting tendencies. The averaged loss curve information as well as the comparison between actual and predicted peel forces of the final model, for the cross validation employed can be found in the supplementary Figure S2.

In fold 1, the training and validation loss curves show rapid convergence during the initial 20 epochs as shown in Figure 5A. The training loss decreases sharply and stabilizes around epoch 40, while the validation loss follows a similar trend, albeit with slightly delayed stabilization with an MSE of $1.48 \times 10^{-6}$. The final loss values of the training and validation sets align closely, indicating minimal overfitting and strong generalizability. The smooth decrease and stabilization of both curves suggest that the data split in this fold facilitated effective learning without introducing significant noise or bias. The loss curves for fold 2 exhibit a slightly stable trend compared to fold 1 (Figure 5B). While the training loss decreases rapidly within the first 50 epochs, the validation loss follows a more gradual decline, converging and stabilizing around epoch 100. The final loss values for the training and validation sets are closely aligned, signifying effective generalization. The gradual reduction in validation loss, coupled with the eventual stabilization, highlights the model's resilience and adaptability when exposed to a diverse subset of the data. The MSE value for this fold was calculated to be $1.06 \times 10^{-7}$. Fold 3 introduces additional complexity, as evidenced by the slightly oscillating training and validation loss curve in Figure 5C. Both training and validation losses decrease rapidly within the first 50 epochs, achieving stabilization by epoch 100. However, the validation loss exhibits minor fluctuations later during the training process, which could be attributed to the characteristics of the validation data subset in this fold. The noise in the validation loss does not result in a significant divergence from the training loss, indicating that the model was able to effectively learn from the training data while maintaining generalizability. The final alignment of the loss values with an MSE of $5.04 \times 10^{-7}$ reinforces the robustness of the model architecture and the chosen hyperparameters. The loss curves for fold 4 demonstrate a steep decline in the early stages of training (from epoch 0 to approximately epoch 30) as shown in Figure 5D, indicating rapid improvement in performance as it learns from the training data. Both the training and validation loss curves exhibit gradual convergence, flattening out beyond epoch 50. The plateau suggests that the model reaches appoint of diminishing returns where additional training results in only marginal improvements. Notably the training and validation losses remain closely aligned throughout the entire training process. This close tracking is indicative of good generalization performance, implying that the model does not overfit. From Figure 5E, the training and validation loss curves of 5$^{th}$ fold demonstrate one of the most stable and consistent performances across all folds. Both curves converge rapidly within the first 50 epochs and remain closely aligned throughout the training process. The minimal gap between the training and validation losses suggests a well-balanced split of the data in this fold, resulting in an optimal training process with negligible overfitting. This fold highlights the model's capability to achieve strong generalization when the training and validation sets share a high degree of similarity in their distributions. The smooth and stable trends in both curves reflect the absence of significant irregularities in the data. Fluctuating behaviour observed in some of the folds underscores the importance of incorporating cross-validation to account for potential variations in data complexity.

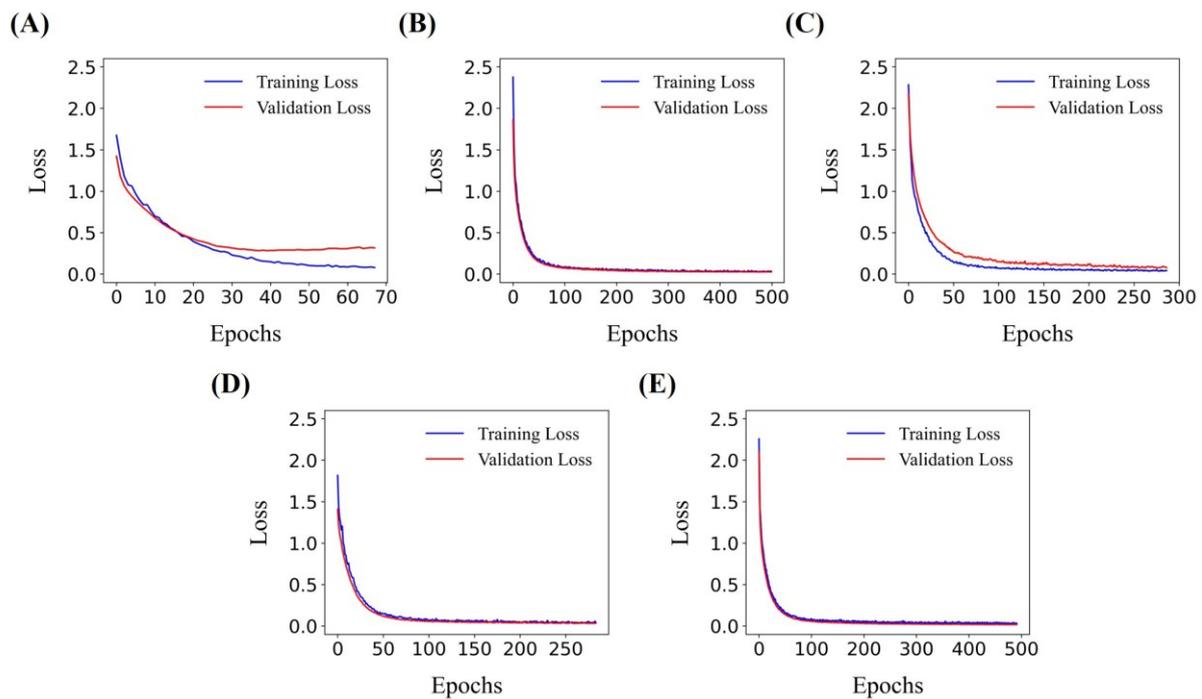

Figure 5: Training and validation loss curves with the number of epochs across each fold of the 5-fold cross validation done on the final model.

Overall, the rapid convergence of training losses in all folds demonstrates the efficiency of the model in learning from the training data. The validation losses, while exhibiting slight variability in convergence rates, consistently align with the training losses in their final stabilized values. This consistent alignment indicates that the model successfully mitigates overfitting, which is often a concern when dealing with small datasets. Folds 2, 4, and 5 demonstrate the most stable loss curves, characterized by minimal fluctuations, MSE values, and higher $R^2$ scores when compared to the remaining folds, indicating these folds likely represent data splits where the training and validation sets were well-balanced, allowing the model to achieve optimal performance with minimal effort. In contrast, folds 1 attained rapid convergence requiring a lesser number of epochs as compared to other folds. The fluctuating validation loss curve in fold 3 reflects the potential influence of outlier data points or increased variability in the validation subset. Despite these differences, the final loss values across all folds are comparable, reinforcing the robustness of the neural network model. The predicted vs actual peel force analysis for each fold of the cross validation is present in the supplementary information Figure S3.

A key advantage of these neural network models is their ability to significantly reduce computational costs when compared to traditional FEM simulations. Training the neural networks with data corresponding to variety of adhesives with varying material and fracture mechanics parameter is completed in just over a minute. Similarly, once the networks are trained, predictions can be generated within minutes, regardless of the quantity. In contrast, FEM simulations are considerably more time-intensive, often requiring approximately 20 to 40 minutes to complete a single simulation. The exact duration depends on various factors, including the complexity of the geometry, the level of meshing required, and the specific parameters involved. Moreover, FEM simulations demand significant expertise in FEM tools and solvers, adding to the overall effort and computational overhead. By integrating FEM models with artificial neural networks, this hybrid approach dramatically reduces computational time, enabling faster analyses once the necessary dataset is prepared. This is particularly advantageous when the data is obtained through experimental methods, as it eliminates the need for repetitive, time-consuming simulations. The ability of neural networks to quickly generalize patterns from existing FEM-generated datasets offers a transformative step in the modeling of adhesive peeling behavior, especially for complex systems like skin-adhesive interactions.

## Conclusion

This study presents a novel approach for predicting the peeling behaviour of adhesives on skin substrate, leveraging neural networks as a computationally efficient alternative to traditional experimental and FEM simulations. By employing adhesive material parameters and fracture mechanics properties as input features, the trained neural network effectively predicts critical force-displacement characteristics with high accuracy. The input and output data were generated using finite element simulations of a 90° peel test, performed by systematically varying the combinations of the adhesive's visco-hyperelastic material properties and cohesive zone parameters. The resulting output for each simulation is the corresponding F-d curve. The minimum peel force (Fmin) required to separate the adhesive from the skin substrate is predicted using the trained neural network. The neural network framework demonstrated remarkable predictive accuracy for the peeling behaviour of adhesives, as evidenced by robust performance metrics across multiple folds of cross-validation of the final model. The two-stage cross-validation approach employed underscores the importance of iterative model evaluation and refinement in neural network development. The first phase of cross-validation enabled the identification of an optimal architecture tailored to the simulation dataset, while the second phase validated the reliability and robustness of the final model. This dual approach ensured the selection of an architecture that balances bias-variance and provided a comprehensive assessment of the model's performance, offering confidence in its predictive capabilities. The main novelties of this work include the proposed framework's ability to predict adhesive peeling behaviour with only material and interfacial parameters as inputs representing a significant advancement in adhesive testing methodologies. Moreover, the proposed neural network framework reduces computational time to mere minutes for training and prediction once the requisite dataset is available. This acceleration in analysis paves the way for broader applicability, particularly in industrial and clinical settings where iterative testing and optimization are critical.

While our models demonstrate excellent performance, its scope and robustness could be further improved by training on larger datasets of adhesive peeling from skin substrates. Future work could be to broaden the scope of this study by incorporating peeling models for various testing angles beyond the 90-degree peel test, enabling a more comprehensive understanding of adhesive behaviour under different geometries. Additionally, the influence of adhesive thickness variation could be systematically investigated to account for real-world applications where thickness plays a critical role. While the present study focused on peeling behaviour at a single interface, multi-layered systems or heterogeneity in the substrates may introduce complexities that require more sophisticated modelling strategies, such as multi-task learning or physics-informed neural networks (PINNs).

In conclusion, our study demonstrates that by leveraging neural networks trained on FEM-generated data, it is possible to achieve high accuracy in predicting peeling behaviour across a range of material and interfacial properties. The trained model offers a fast, reliable, and scalable alternative to traditional FEM simulations, which is particularly valuable in applications requiring iterative design processes or real-time decision-making. Furthermore, this hybrid approach bridges the gap between computational mechanics and data-driven modelling, showcasing the potential of machine learning to enhance the efficiency and accessibility of advanced engineering analyses.

## Data Availability

All the technical details relating to data generation, software used, data processing, model building and evaluation are provided in the manuscript text for easy reproduction of the results. However, the authors are unable to share the full codes and data owing to the strict confidentiality agreements and policies of their employer.

## Supporting Information

Supporting figures consisting of the details about the FEM simulation performed on ANSYS Workbench; peel test geometry and mesh configuration, dataset generation methodology; loss curve of the final model and comparison of the predicted vs actual peel forces for the neural network model.

## Author Contributions

R.G conceived the project idea. N.N.D. assisted A.M. with the FEM simulations and verified the results. A.M. performed the FEM simulations and developed the neural network model. All authors contributed to the interpretation and discussion of the results and preparation of this manuscript.

# Acknowledgements

The authors would like to thank Dr. Harrick Vin, CTO, Tata Consultancy Services and Dr. Sachin Lodha, Head of Research, Tata Consultancy Services for their continuous encouragement and support for this project. The financial support for this research is provided by Tata Consultancy Services, CTO organization.

# Neural networks for the prediction of peel force for skin adhesive interface using FEM simulation

*Ashish Masarkar[1], Rakesh Gupta[1*], Naga Neehar Dingari[1] and Beena Rai[1]*

[1]TCS Research, SP2 Campus, Rajiv Gandhi Infotech Park, Phase 3, Hinjewadi, Pune 411057,

India

*Corresponding Author Email: gupta.rakesh2@tcs.com, n.dingari@tcs.com

## FEM Simulation of 90° Peel Test

In this section, a detailed explanation of the FEM simulation of 90° peel test for generating F-d data is explained.

We used the commercial finite element (FE) software ANSYS Workbench (2023 R2) to model the peel test for tissue adhesive interface. Figure shows the peel test geometry and the boundary conditions used in the FE model. Here in this study a simplified plain-strain condition is assumed. The free end of the peeling arm was subjected to a constant velocity boundary condition as shown in Figure S1 (B), in the direction normal to the interface to mimic the $90^0$-peel test. The bottom surface is constrained in a frictionless manner. Table S1 summarizes the material and element type information for the peel geometry used. For contact modelling Augmented Lagrange formulation was used.

Mesh convergence study determined that an element size of 0.5 mm was optimum to ensure numerical accuracy and solution stability. The final discretized model consisted of 4016 nodes and 1120 elements. The simulations were executed using an implicit solver to ensure numerical stability under quasi-static conditions. The reaction forces at the free end of peel arm and corresponding displacement values were recorded to construct F-d curves.

Table S4 Material and element type used for FEM simulation in the ANSYS Workbench

| Element Name | Element Type | Element Shape | No. of Elements |
|---|---|---|---|
| Adhesive and Tissue | PLANE 183 | Quad8 | 1120 |
| Interface Contact Body | CONTAC172 | LINE3 | 160 |
| Interface Target Body | TARGE 169 | LINE3 | 160 |

**Dataset Generation**

Table S2 summarizes the range of the visco-hyperelastic material parameters for the adhesive and the CZM parameters for the interface for generating the F-d dataset. The relaxation times ($t_1$-$t_5$) were varied across different time scales, ranging from $10^{-4}$ s to 500 s. The times were selected to span multiple orders of magnitude, ensuring that both short-term (instantaneous response) and long-term (gradual relaxation) effects were captured. Some configurations used a logarithmic distribution of times (e.g., $10^{-4}, 10^{-2}, 1, 10, 100$), while others followed non-uniform increments (e.g., $10^{-3}, 10^{-1}, 10, 50, 200$). The lower values of $t_i$ (e.g., $10^{-4}$, $10^{-3}$ s) correspond to fast-relaxing components, while higher values (e.g., 100, 500 s) represent long-term stress relaxation effects. Similarly, the moduli values ($G_1$-$G_5$) were varied between 0.03-0.13 allowing for a range of adhesive stiffness values. Higher values (0.12–0.13) correspond to adhesives that retain more stress over time, leading to a stiffer viscoelastic response. Lower values (~0.03–0.05) indicate a more compliant adhesive, where relaxation occurs faster, and stress dissipates quickly. The incremental variation across different datasets ensures a diverse range of viscoelastic properties, capturing behaviours from highly elastic to highly viscous adhesives.

By systematically varying both relaxation times and moduli, the dataset effectively captures a wide spectrum of adhesive behaviour:

- Short term response (low $t_i$ and high $G_i$): Adhesives with rapid stress dissipation, suitable for soft adhesion.
- Long-term response (high $t_i$ and low $G_i$): Adhesives with extended stress retention, mimicking biomedical or structural adhesives.

For Ogden's material model the shear modulus µ, which defines the material's resistance to shear deformation was varied from 0.5 MPa to 1.4 MPa in increments of 0.1 MPa. This range was selected to encompass soft adhesive materials while ensuring computational stability in the FEM simulations. The parameter α governs the degree of nonlinearity in the stress-strain response and was varied between 2.0 – 2.9, increasing by 0.1 per step. A higher value results in a more pronounced nonlinear behavior, which influences the F-d response of the adhesive during peeling. The incompressibility factor is calculated by the following relation:

$$D = \frac{1}{2\mu}$$

As the shear modulus increased, D exhibited a decreasing trend, ranging from 2.0 to 0.714. The systematic variation of µ, α and D enabled the exploration of different adhesive mechanical behaviors, ranging from softer, highly deformable adhesives to stiffer materials with reduced compliance. The F-d responses obtained from these parameter combinations were used to analyze the influence of hyperelastic properties on peel force and adhesion strength. The selected range ensured that the generated force-displacement curves encompassed realistic adhesive behaviours while maintaining convergence in the finite element simulations.

The fracture mechanics parameters, particularly $\sigma_{max}$ was varied between 1 kPa and 10 kPa (0.001 MPa to 0.01 MPa) representing the maximum tensile stress the adhesive interface can withstand before failure occurs. Lower values (1-3 kPa) indicate weak adhesion strength where the adhesive detaches under minimal force, whereas higher values (9-10 kPa) corresponds to stronger adhesion, requiring a greater force to initiate debonding. The incremental variation (0.001 MPa per step) ensures a smooth gradient of adhesion strengths, allowing the model to capture subtle differences in peeling behaviour across different adhesive formulations. The fracture energy '$G$' was varied between 23 J/m² and 32 J/m². The stepwise increase (1 J/m² per step) allows for fine-tuned variations, enabling an in-depth study of gradual changes in adhesive performance. These variations ensured a diverse and representative dataset, enabling the development of a robust predictive model for analysing adhesive peeling behaviour on skin.

Table S5 Range of the visco-hyperelastic material model for the adhesive and CZM parameters for generating the F-D dataset.

| Prony Series Parameter | | Ogden's Parameter | | CZM Parameter | |
|---|---|---|---|---|---|
| $t_i$ | $G_i$ | $\mu_{adhesive}$ (MPa) | $\alpha_{adhesive}$ | $G$ (J/m²) | $\sigma_{max}$ (KPa) |
| 1e-4 – 5e+2 | 0.03 – 0.13 | 0.5 – 1.4 | 2 – 2.9 | 23 – 32 | 1 - 10 |

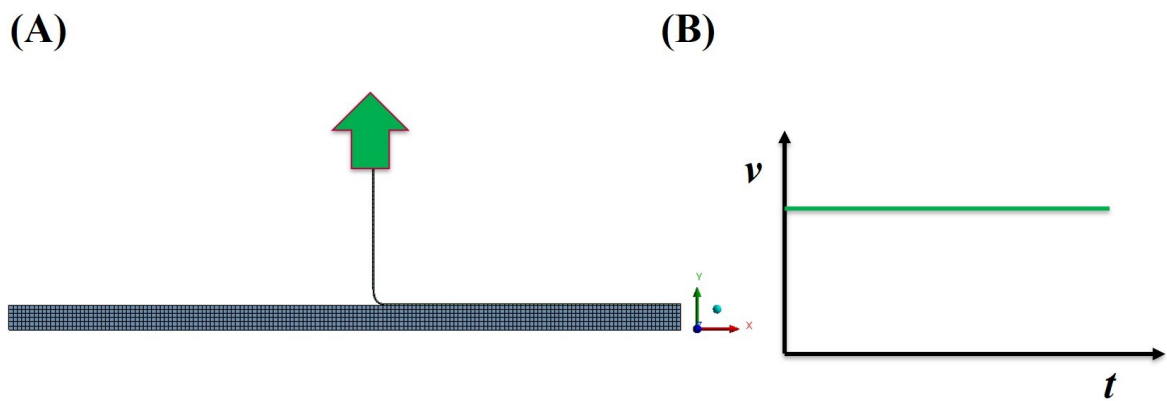

Figure S1 (A) Geometrical and boundary conditions in the FEM model with mesh element size of 0.5 mm (B) Velocity (5mm/sec) boundary condition applied to the free end of the peeling arm

## Background theory on Neural Networks

A feedforward artificial neural network (ANN) is a computational model composed of layers of interconnected nodes or "neurons," inspired by the human brain. It maps input features to an output through a series of transformations, enabling it to learn complex nonlinear relationships. The ANN consists of:

- **Input Layer:** Receives the input features $x = [x_1, x_2 \ldots \ldots x_n]^T$
- **Hidden Layer(s):** One or more layers where nonlinear transformations are applied.
- **Output Layer:** Produces the final prediction $\hat{y}$

Each neuron in a layer receives input from the previous layer and processes it using a weighted sum followed by an activation function. For a given neuron $j$ in layer $l$, the weighted input is:

$$z_j^{(l)} = \sum_{i=1}^{n^{(l-1)}} w_{ji}^{(l)} a_i^{(l-1)} + b_j^{(l)} \tag{1}$$

Where:

- $w_{ji}^{(l)}$ = weight from neuron *i* in layer *l-1* to neuron j in layer *l*
- $b_j^{(l)}$ = bias of neuron *j* in layer *l*
- $a_i^{(l-1)}$ = activation from neuron *i* in the previous layer
- $z_j^{(l)}$ = input to activation function for neuron *j*

The activation function introduces nonlinearity:

$$a_j^{(l)} = f\left(z_j^{(l)}\right) \tag{2}$$

For hidden layers, the ReLU (Rectified Linear Unit) function is commonly used:

$$f(z) = ReLU(z) = \max(0, z) \tag{3}$$

For regression problems, the output layer typically uses a linear activation:

$$\hat{y} = z^{output} \tag{4}$$

To measure the difference between predicted and actual values, MSE loss is used:

$$L(\hat{y}, y) = \frac{1}{N} \sum_{i=1}^{N} (\hat{y}_i - y_i)^2 \tag{5}$$

Where N is the number of samples, $\hat{y}_i$ is the predicted output and $y_i$ is the true value. To minimize the loss, gradients of the loss with respect to weights are computed using backpropagation, and weights are updated via gradient descent. Weight update rule using the Adam optimizer is given by:

$$\theta_{t+1} = \theta_t - \eta \cdot \frac{m_t}{\sqrt{v_t} + \epsilon} \tag{6}$$

Where:

$\theta_t$ = model parameter at step *t*, $\eta$ = learning rate, $m_t, v_t$ = first and second moment estimates of gradients and $\epsilon$ = small value for numerical stability.

To reduce overfitting L2 regularization (Ridge) adds a penalty term to the loss as:

$$L_{reg} = L + \lambda \sum_{j,l} \left(w_{ji}^{(l)}\right)^2 \tag{7}$$

Where λ is the regularization coefficient.

The dropout layer randomly drops a fraction $p$ of neurons during training:

$$a_j^{(l)} = \begin{cases} 0 & \text{with probability } p \\ \dfrac{a_j^{(l)}}{1-p} & \text{otherwise} \end{cases} \qquad (8)$$

Batch normalization normalizes activations to stabilize training:

$$\hat{z}_j = \frac{z_j - \mu B}{\sqrt{\sigma_B^2 + \epsilon}}, a_j = \Upsilon \hat{z}_j + \beta$$

Where $\mu B$ and $\sigma_B^2$ are the batch mean and variance.

Table S6 Summary of the hyperparameters used for the final neural network model

| Hyperparameter | Value |
| --- | --- |
| Hidden Layers | 1 |
| Neurons per hidden layer | 32 |
| Activation function | ReLU |
| Output activation | Linear |
| L2 Regularization (λ) | 0.05 |
| Dropout rate | 0.4 |
| Batch normalization | Yes |
| Learning rate | 0.001 |
| Maximum epochs | 2000 |
| Learning rate scheduler | ReduceLROnPlateau (factor = 0.9, patience = 10, min_lr = 1e-6) |
| Early stopping patience | 20 |
| Seed | 19 |

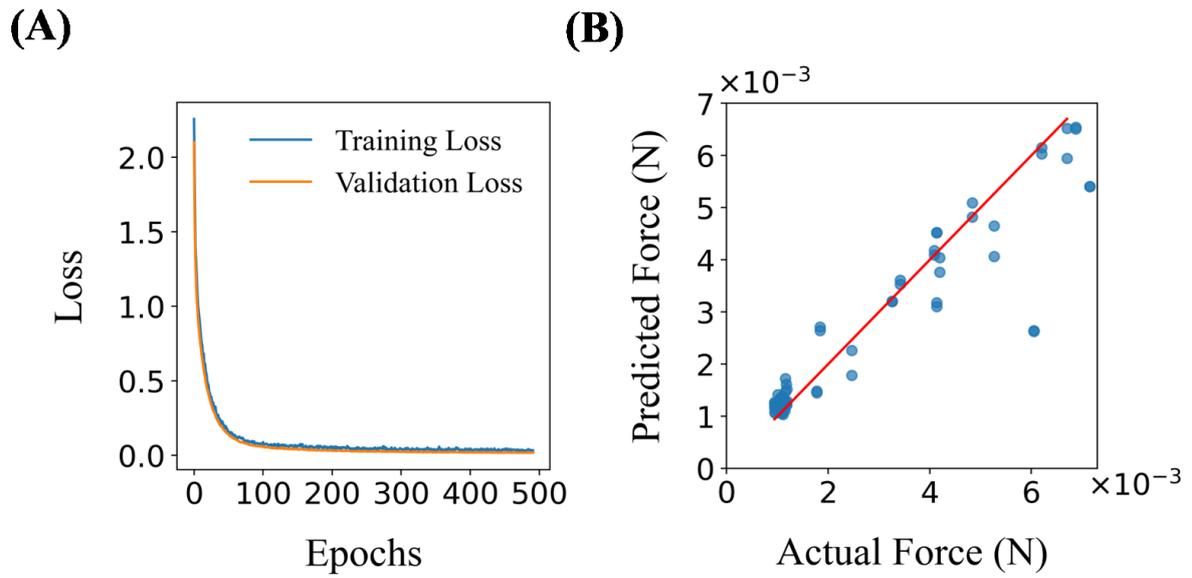

Figure S2 (A) Averaged training and validation loss curve (B) Averaged actual vs predicted peel forces of the 5-fold cross validation employed on the final model

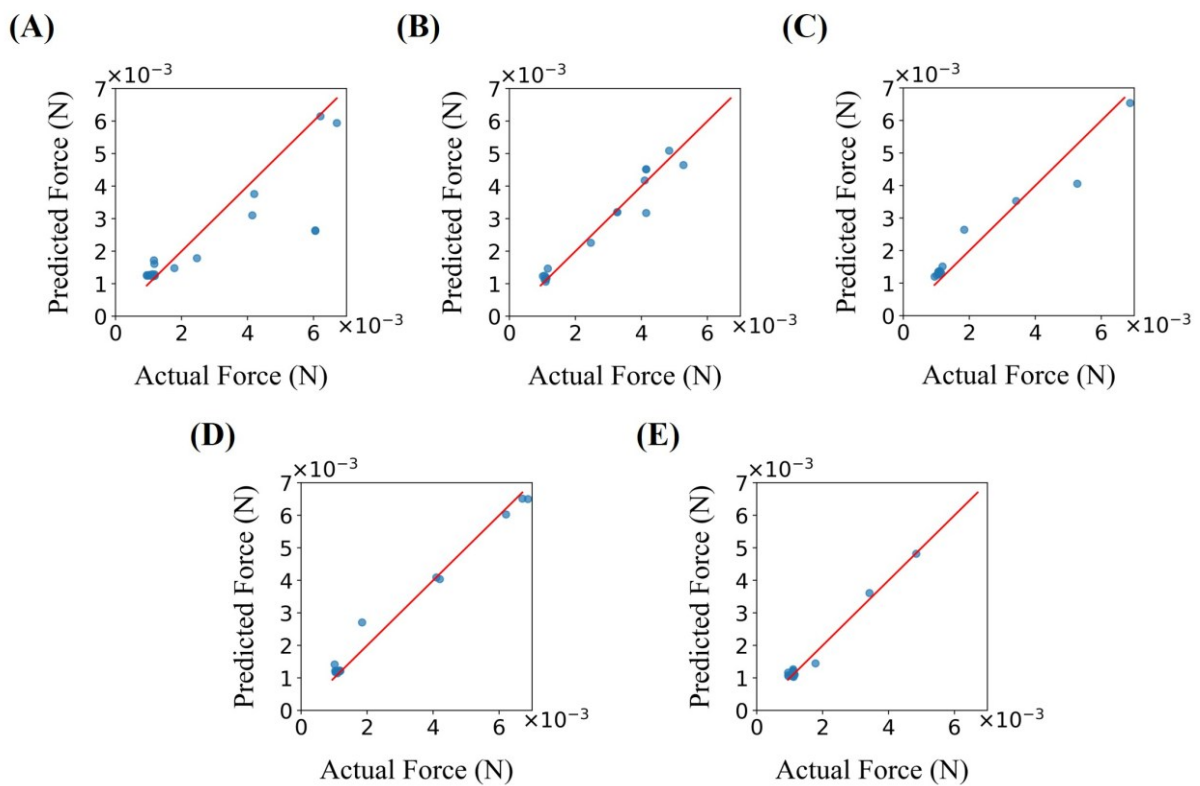

Figure S3 Actual vs predicted peel forces across the 5 folds (A)-(E) of the final model